\documentstyle[aps,epsfig,multicol]{revtex}
\begin{document}
\begin{flushright}
KIAS Preprint P02009 \\
OCHA-PP-188 \\
KEK-TH-813 \\
May 2002 \\
\end{flushright}
\vspace{1cm}
\baselineskip 24pt 
\begin{center}
{\Large\sc {\bf }}
\vspace*{3mm}
{\Large\sc {\bf On the possibility of a very light $A^0$ 
at low $\tan\beta$ \\
in the MSSM}} 
\end{center}
\vspace{1cm}
\baselineskip 18pt 

\begin{center}
{\large
{{A.G. Akeroyd}$^{\rm a,}$\footnote{E--mail: akeroyd@kias.re.kr},
{S. Baek}$^{\rm a,}$\footnote{E--mail: swbaek@kias.re.kr}, 
{Gi-Chol Cho}$^{\rm b,}$\footnote{E--mail: cho@phys.ocha.ac.jp},
K. Hagiwara$^{\rm c,}$\footnote{E--mail: kaoru@post.kek.jp}
}}
\vspace{.4cm}
{\it
\\
a: Korea Institute for Advanced Study, 207-43 Cheongryangri-dong,\\
Dongdaemun-gu, Seoul 130-012, Republic of Korea\\

\vspace{.4cm}
b: Department of Physics, Ochanomizu University,\\ 
Tokyo 112-8610, Japan\\

\vspace{.4cm}
c: KEK Theory Group, 1-1 Oho \\
Tsukuba, Ibaraki 305-0801, Japan

}

\end{center}

\vspace{1cm}
\begin{abstract}
\noindent
The searches at LEP II for the processes $e^+e^-\to h^0Z$ and
$e^+e^-\to h^0A^0$ in the Minimal Supersymmetric Standard Model (MSSM)
fail to exclude regions of the $m_h,m_A$ plane
where $\tan\beta <1$, thus allowing a very light $A^0$
($m_A< 20$ GeV). Such a parameter choice would predict a 
light $H^\pm$ with $m_{H^\pm}< m_W$. Although
the potentially large branching ratio for $H^\pm \to A^0 W^*$
would ensure that $H^\pm$ also escaped detection in direct searches
at LEP II and the Tevatron Run I, we show that 
this elusive parameter space is overwhelmingly 
disfavoured by electroweak precision measurements through
its large contribution to the $Zb\overline b$ vertex.

\end{abstract}

\def\thefootnote{\alph{footnote}}
\setcounter{footnote}{0}
\newpage
\section{Introduction}
The Minimal Supersymmetric Standard Model (MSSM) is currently
the leading candidate for physics beyond the Standard Model (SM). 
The MSSM predicts five physical Higgs bosons: a charged pair ($H^+$,$H^-$),
two CP--even scalars ($h^0$,$H^0$), and a CP--odd pseudoscalar $A^0$.
To date their detection remains elusive despite intense 
searches at past and present colliders.  
Lower limits on their masses have been derived, which assume
certain decay modes and apply to specific regions of the 
MSSM parameter space \cite{Groom:in}.
LEP II has the strongest lower limits on the masses of all the above
Higgs bosons. In the case of $H^\pm$, limits from the Tevatron run I can 
be competitive in specific regions of parameter space 
(small and large $\tan\beta$)\cite{Abbott:1999ec,Affolder:1999au}.
LEP carries out searches in the channels $e^+e^-\to h^0Z,h^0A^0$ and 
obtains the mass limits $m_A> 91$ GeV, $m_h> 91$ GeV, 
for $\tan\beta> 1$ \cite{unknown:2001xx}. In the framework
of the MSSM, these mass limits enable indirect bounds on $m_{H^\pm}$ and 
$m_{H^0}$ to be derived which are stronger than the direct search limits
for these particles \cite{Drees:1998pw}.
For example, $m_A> 90$ GeV implies $m_{H^\pm}> 120$ GeV, 
which is stronger than the direct bound $m_{H^\pm}> 78$ GeV.

The above mass bounds are obtained for $\tan\beta> 1$.
In this paper we concentrate on the region of $\tan\beta < 1$
in which the mass bounds are considerably weakened. 
In particular, a very light $A^0$ ($m_A< 20$ GeV) has not been 
excluded by direct searches \cite{unknown:2001xx}.
The MSSM with $\tan\beta < 1$ is  not strictly the minimal model since 
it calls for a new physics scale below the GUT scale where the 
top--quark Yukawa coupling becomes strongly interacting.  
We nevertheless advocate consideration of the
$\tan\beta<1 $ parameter space for the following reasons:
\begin{itemize}

\item[{(i)}] Theoretical studies and searches for the Higgs bosons of the 
general two--Higgs doublet model consider $\tan\beta< 1$ 
\cite{Abbiendi:2000ug}. In such models one usually imposes the milder 
requirement that the top--quark Yukawa coupling is perturbative at 
the weak scale, corresponding to $\tan\beta> 0.4$ \cite{Haber:1997dt}.

\item[{(ii)}] A very light pseudoscalar is a necessary condition for 
spontaneous CP violation in the MSSM 
\cite{Maekawa:1992un,Pomarol:1992bm,Kong:1997ux,Lebedev:fw}
i.e.  the existence of a relative
phase between the vacuum expectation values of the two Higgs doublets.

\end{itemize}
Point (ii) is of particular interest since spontaneous CP violation
has been considered ruled out in the MSSM for years due to the 
progressively stronger mass bound on $m_A$ for $\tan\beta> 1$. 
However, spontaneous CP violation might be possible
in the unexcluded region of low $\tan\beta$ and light $m_A$
(despite a possible new physics threshold much below the GUT scale), thus
motivating experimental consideration of this region.

A very light $A^0$ is not ruled out for $\tan\beta< 1$ because the 
standard search strategies which rely on $b$ tagging the decays
$h^0\to b\overline b$ or $h^0\to A^0A^0 \to b\overline b b \overline b$ 
become ineffective. In the $\tan\beta < 1$
region the decay $h^0\to A^0A^0$ becomes increasingly important
\cite{Brignole:1991pq},
thus suppressing BR$(h^0\to b\overline b$). The subsequent decay  
$A^0\to c\overline c$ has a sizeable branching ratio (BR) 
for $\tan\beta< 1$ and/or $m_A> 2m_b$
and so suppresses BR$(A^0\to b\overline b$), giving 
rise to topologies of $c\overline c c\overline c c\overline c$ (from 
$e^+e^-\to h^0A^0$) or $Z c\overline c c\overline c$
(from $e^+e^-\to h^0Z)$. These final states elude the standard LEP searches,
resulting in unexcluded regions in the ($m_h,m_A$) plane  
\cite{unknown:2001xx}. 

The existence of a very light $A^0$ in the MSSM requires a light $H^\pm$ 
with $m_{H^\pm}< $ 80 GeV, which may violate constraints on
$m_{H^\pm}$ from other experiments e.g. i) direct
searches at LEP II and the Tevatron Run I, 
and ii) the indirect effect of $H^\pm$ on flavour changing
neutral current processes (FCNC) and electroweak 
precision measurements. Thus in the framework of the MSSM the
unexcluded region of very light $A^0$ may be probed by searching for
a light $H^\pm$. We will show that the direct searches for 
$H^\pm$ would be affected
by the presence of a large BR for $H^\pm\to A^0W^*$, and therefore
such a $H^\pm$ may not be ruled out from current searches
\footnote{This was first pointed out for LEP II searches in 
\cite{unknown:2001xx}.}.
However, the indirect effect of $H^\pm$ on electroweak 
observables which are sensitive to corrections to the 
$Zb_Lb_L$ vertex is very pronounced,
and a light $H^\pm$ with $\tan\beta< 1$ is strongly disfavoured by 
LEP electroweak precision data.

Our work is organized as follows. In Section 2 we evaluate the 
BR($H^\pm\to A^0W^*$) in the region of ($m_A,\tan\beta$) 
unexcluded by direct searches.
Section 3 considers the indirect effect of a light $H^\pm$
at low $\tan\beta$ on electroweak precision observables, while section 4 
contains our conclusions.

\section{Direct searches for $H^\pm$}

The neutral Higgs boson searches at LEP in the channels
$e^+e^-\to Zh^0$ and $e^+e^-\to h^0A^0$ exclude regions in the 
($m_h,m_A$) plane. There are three benchmark scenarios used in the 
LEP searches:

\begin{itemize}

\item [{(i)}] No mixing case: defined by $X_t=A_t-\mu\cot\beta=0$

\item [{(ii)}] Maximum $m_h$ case: defined by $X_t=2M_{SUSY}$

\item [{(iii)}] Large $\mu$ case

\end{itemize}

In these scenarios all SUSY parameters are fixed.
In scenario (i) an unexcluded region in the 
($m_h,m_A$) plane still remains for $m_A< 20$ GeV and 
$65$ GeV$< m_h < 80$ GeV, 
provided $\tan\beta< 0.8$. This corresponds to the region where the 
decay chain $h^0\to A^0A^0\to c\overline c c\overline c$ is important.


In the following subsections we shall study the phenomenology
of $H^\pm$ corresponding to the region unexcluded by direct searches for
$h^0$ and $A^0$, which predicts 
$m_{H^\pm} < 76(78)$ GeV for $\tan\beta=0.8(0.6)$. 
It should be noted that there are sizeable corrections to the
tree level sum rule, $m_{H^\pm}=\sqrt{m_W^2+m_A^2}$~\cite{Diaz:1991ki},
and $m_{H^\pm}$ can be as light as 69 (73) GeV for $\tan\beta=0.6 (0.8)$.
In this region there exists the following mass hierarchy,
$m_{H^0} \gg m_{h^0} \approx m_{H^\pm} \gg m_{A^0}$ which differs 
from the usual
Higgs mass hierarchy $m_{H^\pm} \approx m_{H^0} \approx m_{A^0} > m_{h^0}$ 
for $\tan\beta>1$ and $m_{A^0} > 130$ GeV.
We will show results for the no mixing case, 
using the LEP values $\mu=-200$ GeV, $M_{SUSY}=1$ TeV.

\subsection{The decay $H^\pm\to A^0W^*$}

In the parameter space of low $\tan\beta$ and small $m_A$
the three body decay $H^\pm\to A^0W^*\to A^0ff'$  
is sizeable. In Fig. 1 we show
BR$(H^\pm\to A^0W^*$) as a function of $m_{H^\pm}$.
One observes that BR$(H^\pm\to A^0W^*)$ can be as large as 60\% (45\%)
for $\tan\beta=0.8(0.6)$. Note that these BRs are larger than
those presented in \cite{Djouadi:1995gv} since this paper used
$\tan\beta=1.6$ and the (then) bound $m_A> 50$ GeV. 

\subsection{$H^\pm$ searches at LEP II}

LEP II searches for $e^+e^-\to H^+H^-$ and has derived limits on $m_{H^\pm}$  
as a function of BR$(H^\pm\to \tau\nu_\tau$), with the most
conservative limit being $M_{H^\pm} > 78.6$ GeV  \cite{unknown:2001xy}.
The searches assume
BR$(H^\pm\to \tau\nu_\tau)$+BR$(H^\pm\to cs)=1$.
The large BR$(H^\pm\to A^0W^*)$ in
the parameter space of interest would
weaken the existing limits on $m_{H^\pm}$, allowing $m_{H^\pm} 
< 76(78)$ GeV for $\tan\beta=0.8(0.6)$ i.e. the prediction for
$m_{H^\pm}$ in the unexcluded region. Investigation of the 
impact of $H^\pm\to A^0W^*\to Aff'$ decays on the MSSM $H^\pm$ searches
at LEP is in progress, and a preliminary search 
\cite{OPAL:HAW} for such decays
has been performed in the context of the 2HDM (Model I), where
BR($H^\pm\to A^0W^*$) can reach $100\%$ \cite{Akeroyd:1998dt}.
However, this search also required $b$--tagging and is only valid 
for $m_A > 2 m_b$ and large BR$(A^0 \to b\overline b)$.


\subsection{$H^\pm$ searches at the Tevatron}

The Tevatron Run I searched for $t\to H^\pm b$ decays and rules out
regions of the ($m_{H^\pm},\tan\beta$) plane which would correspond to 
values of BR$(t\to H^\pm b$) incompatible with the observed events in 
$t \overline t$ production. The exact boundary of the excluded
region depends sensitively on the theoretical cross section,
$\sigma(p \overline{p} \to t\overline t$), with the most 
conservative excluded region occurring for larger theoretical
cross--section. The region $m_{H^\pm}< 80$ GeV and
$\tan\beta< 0.8$ has been ruled out at $95\%$ CL
\cite{Abbott:1999ec}.

In the low $\tan\beta$ region, where $H^+ \to c\overline s$
is normally assumed to be the dominant channel, 
the contribution of $t \to b H^+ \to b c \overline s$ decays
would cause a deficit in the expected number of 
leptonic ($e,\mu$) decays of the top ($t\to bW^+\to bl^+ \nu_l$), 
thus giving a worse fit to the data.
The suppression of BR($H^\pm \to cs,\tau\nu_{\tau})$
by $H^\pm \to A^0W^*$ (shown in Fig. 1)
would be sufficient to shift the boundary of the excluded region to 
lower values of $\tan\beta$, thus narrowly permitting $m_{H^\pm}< m_W$
and $0.6<\tan\beta<0.8$.
The new decay channel can give rise to the observed final states via 
$H^\pm \to A^0W^* \to A^0 l^\pm\nu_l$.
This was pointed out in \cite{Borzumati:1998xr} and \cite{Ma:1997up}
which concentrated on the region $\tan\beta> 1$ and $m_A> 90$ GeV.
With 2 fb$^-1$ of data at the Tevatron Run II, 
the region $m_{H^\pm}< 80$ GeV and
$\tan\beta< 2.5$ will be ruled out at $95\%$ CL\cite{Carena:2000yx}.
Therefore this scenario of very light $A^0,h^0$ and light $H^\pm$ 
will be excluded or confirmed directly in $t\to H^\pm b$ decays very soon.

\section{Indirect searches for $H^\pm$}

The previous section showed that the unexcluded region from
the searches for $e^+e^-\to h^0A^0$ and $e^+e^-\to Zh^0$ 
corresponds to a light $H^\pm$ which itself would have
narrowly escaped detection in direct searches due to
the large BR$(H^\pm\to A^0W^*)$. It is known that a light $H^\pm$
and low $\tan\beta$ would have large effects on various rare $K$ 
or $B$ meson decays as well as the electroweak precision measurements. 
In the following subsections we address the
compatibility of $m_{H^\pm}< m_W$ and $\tan\beta < 1 $
with these indirect constraints.

\subsection{Constraints from FCNC processes} 

Owing to the large top quark Yukawa coupling, a light $H^\pm$ 
with $\tan\beta <1$ would contribute strongly to processes like 
$b\to s\gamma$, $B^0\overline {B^0}$ and $K^0\overline {K^0}$ 
mixing, through top quark exchange 
diagrams \cite{Bertolini:1990if,Branco:cj}. 
However, contributions to these processes from supersymmetric particles 
depend strongly on the flavour structure of the MSSM which is 
completely arbitrary. It is known that the $H^\pm$ 
contribution to $b \to s\gamma$ can interfere destructively with the 
chargino contribution depending on the sign of $A_t$, $\mu$ and
the mass insertion parameter $\delta^u_{23}$ in the up squark sector,
\cite{Bertolini:1990if,Oshimo:zd,Misiak:1997ei}.
More importantly, the unconstrained MSSM allows for
gluino mediated $b-s$ transitions with arbitrary 
coupling strength $\delta^d_{23}$ \cite{Gabbiani:1996hi},
which can be chosen to cancel any large contribution from $H^\pm$.
Given the possibility of FCNCs mediated by supersymmetric particles 
we conclude that no definite constraints on $H^\pm$ 
can be derived from these processes. 

\subsection{Constraints from electroweak precision measurements}

Stringent constraints on a light $H^\pm$ are obtained from 
electroweak precision measurements, i.e. the $Z$-pole observables 
from LEP1 and SLC, and the $W$ mass from Tevatron and LEP2. 
The $Z$-pole observables consist of 8 line-shape parameters 
$\Gamma_Z, \sigma_h^0, R_l, A_{\rm FB}^{0,l}(l=e,\mu,\tau)$,  
two asymmetries from the $\tau$-polarization data ($A_\tau$, 
$A_e$), the decay rates and the asymmetries of $b$- and 
$c$-quarks ($R_b,R_c,A_{\rm FB}^{0,b},A_{\rm FB}^{0,c}$) 
and the asymmetries measured at SLC $(A_{\rm LR}^0,A_b,A_c)$.  
We use the experimental data of these observables 
reported at the 2001 summer conference\cite{lepton-photon}. 
Taking account of the $m_t$ data from Tevatron\cite{mtop}, 
the QCD coupling $\alpha_s(m_Z)$\cite{Groom:in} and the QED 
coupling $\alpha(m_Z^2)$\cite{BP01} at the $Z^0$-mass scale, 
we find that the SM best fit gives 
$\chi^2/({\rm d.o.f.})=25.3/(21-4)$ (9\% CL) \cite{CH2000ew}. 

The Higgs bosons in the MSSM affect these electroweak observables 
through the radiative corrections to the gauge boson two-point 
functions (oblique corrections), and the vertex and box corrections. 
Among them, the $Z b_L b_L$ vertex is most sensitive to small 
$m_A$ (i.e., small $m_{H^\pm}$) and small $\tan\beta$ through the 
charged Higgs boson and the top quark exchange, since the coupling 
$b_L$-$t_R$-$H^\pm$ is proportional to $m_t/\tan\beta$. 
In Fig. 2 we show the total $\chi^2$ as a function of $m_A$ for 
various values of $\tan\beta$. 
Four thick lines correspond to $\tan\beta=0.5$, 0.6, 
0.7 and 1.
The thick-dashed horizontal line denotes the minimum $\chi^2$ 
in the SM ($\chi^2=25.3$). 
The experimentally unexcluded $A^0$ mass is $m_A < 40$ GeV 
for $\tan\beta < 0.7$ \cite{unknown:2001xx}
(for the no mixing scenario that we consider). 

It is easy to see that small $\tan\beta (< 1)$ leads to 
a significantly large $\chi^2$ ($\sim 100$ for $m_A < 100$ GeV) 
due to the enhancement of the $b_L$-$t_R$-$H^\pm$ coupling which 
sizeably affects the partial decay width of $Z\to b\bar{b} (\Gamma_b)$. 
The observables which are sensitive to $\Gamma_b$ are $\Gamma_Z$, 
$R_l$ and $R_b$. 
For example, for $\tan\beta=0.7$ and $m_A=20$ GeV which corresponds 
to $m_{H^\pm} \approx 74$ GeV, the deviations of these three observables 
from the experimental data are $3.4\sigma(\Gamma_Z)$, $4.7\sigma(R_l)$ 
and $6.3\sigma(R_b)$, while the SM best fit gives $-0.6\sigma$, 
$0.9\sigma$ and $0.8\sigma$, respectively. 
The total $\chi^2$ for $\tan\beta < 1$ is almost flat for 
$m_A < 30$~GeV. 
This is because only $H^\pm$ sizeably contributes to the 
electroweak observables for $\tan\beta < 1$, and $m_{H^\pm}$ 
varies little with $m_A$ in this range. 
Since the fit shows 
$\Delta \chi^2 \equiv \chi^2_{\rm tot}({\rm MSSM})-\chi^2_{\rm tot}({\rm SM}) 
\approx 75$ for small $m_A$ and small $\tan\beta$, the possibility 
of a very light $A^0$ is strongly disfavoured from electroweak 
precision measurements. 

In the context of a non-supersymmetric two Higgs doublet model
(2HDM) we note that a very light $A^0$ (even $m_A=0.2$ GeV) 
is still consistent with electroweak precision data 
\cite{Larios:2002ha}. This is because the masses of the 
Higgs bosons in the 2HDM are free parameters 
(in contrast to the MSSM) and thus a very light $A^0$
does not require the other Higgs bosons to be light.


\section{Conclusions}
The direct searches at LEP II for the neutral Higgs bosons $A^0$ and $h^0$ 
of the MSSM fail to exclude a light $A^0$ in the region of $\tan\beta < 1$.
The mass of $H^\pm$ is predicted to be $\le 76(78)$ GeV for 
$\tan\beta=0.8(0.6)$, with a large branching ratio for $H^\pm \to A^0W^*$. 
The latter decay ensures that $H^\pm$ would
have escaped the LEP searches, which rely on $H^\pm\to cs (\tau\nu)$ decays.
Such a $H^\pm$ would have also have (narrowly) escaped detection at 
Run I of the
Tevatron in the channel $t\to H^\pm b$, which is very sensitive to the region
$m_{H^\pm}< 80$ GeV and $\tan\beta< 1$. 
The large BR$(H^\pm\to A^0W^*$) shifts the excluded region to lower values of
$\tan\beta$, thus allowing $H^\pm$ in the parameter space of interest.
Run II will comfortably 
detect or rule out the region of $0.6<\tan\beta< 0.8$ and 
$m_{H^\pm}< 78$ GeV, which in turn would directly confirm or rule out the 
possibility of a light $A^0$ ($m_A < 20$ GeV).

Such a $H^\pm$ would sizeably affect several low energy processes 
($B^0,K^0$ decays) and electroweak precision measurements at LEP1 
and SLC. 
Since the former are very sensitive to the (unknown) flavour structure
of the MSSM, no definite constraints on $H^\pm$ can be derived. However,    
we showed that $m_{H^\pm}<m_W$ and $\tan\beta< 1$ 
is strongly disfavoured by electroweak precision data, 
giving rise to exceedingly large $\chi^2$ values. Thus we conclude that
the region of very light $A^0$ and $\tan\beta< 1$, although
unexcluded by direct searches, is severely disfavoured by indirect 
constraints and will be probed directly in $t\to H^\pm b$ decays at the 
Tevatron Run II.

\section*{Acknowledgements}
We thank F. Borzumati, T. Junk and C.P. Yuan for useful discussions.



\begin{figure}[t]
\smallskip\smallskip  
\centerline{{\hskip0.352cm\epsfxsize6 in \epsffile{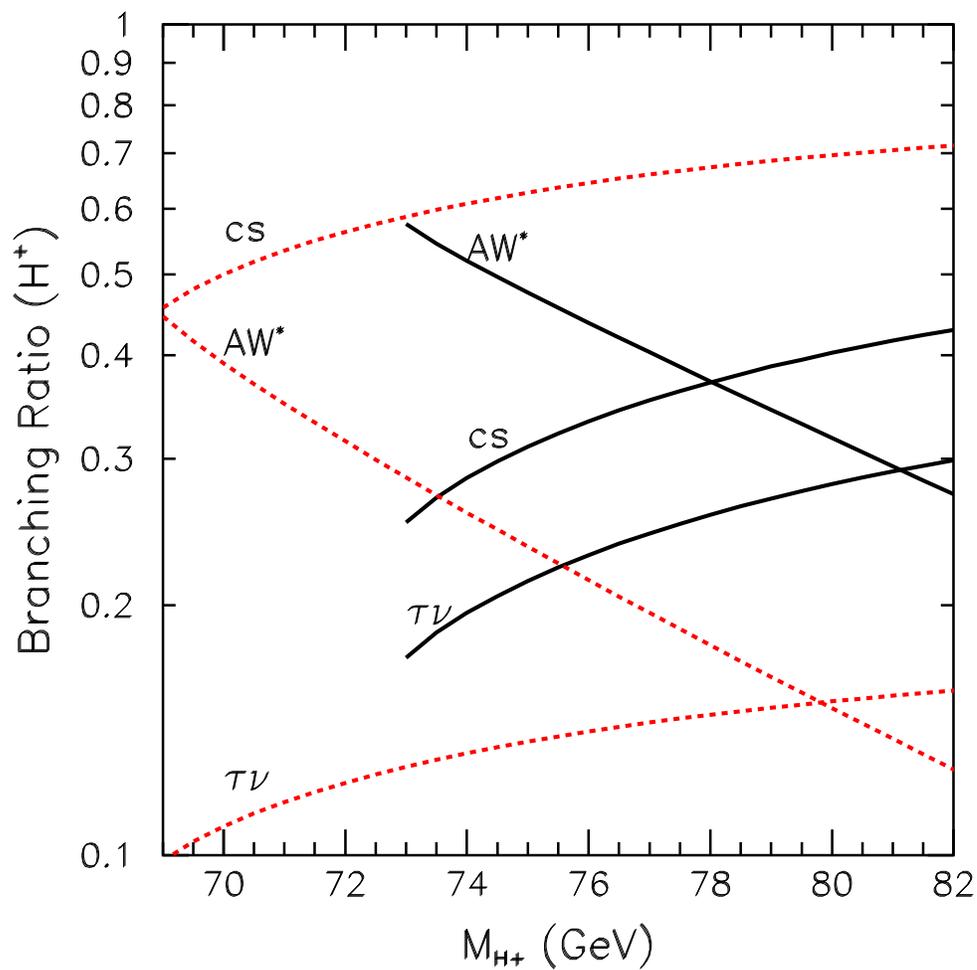}}
}
\smallskip\smallskip\smallskip\smallskip
\caption{Branching ratios of $H^\pm$ as a function of
$m_{H^+}$ for $\tan\beta=0.6$ (dashed lines) and 0.8 (solid lines).
}
\end{figure}

\begin{figure}[t]
\smallskip\smallskip  
\centerline{{\hskip0.352cm\epsfxsize6 in \epsffile{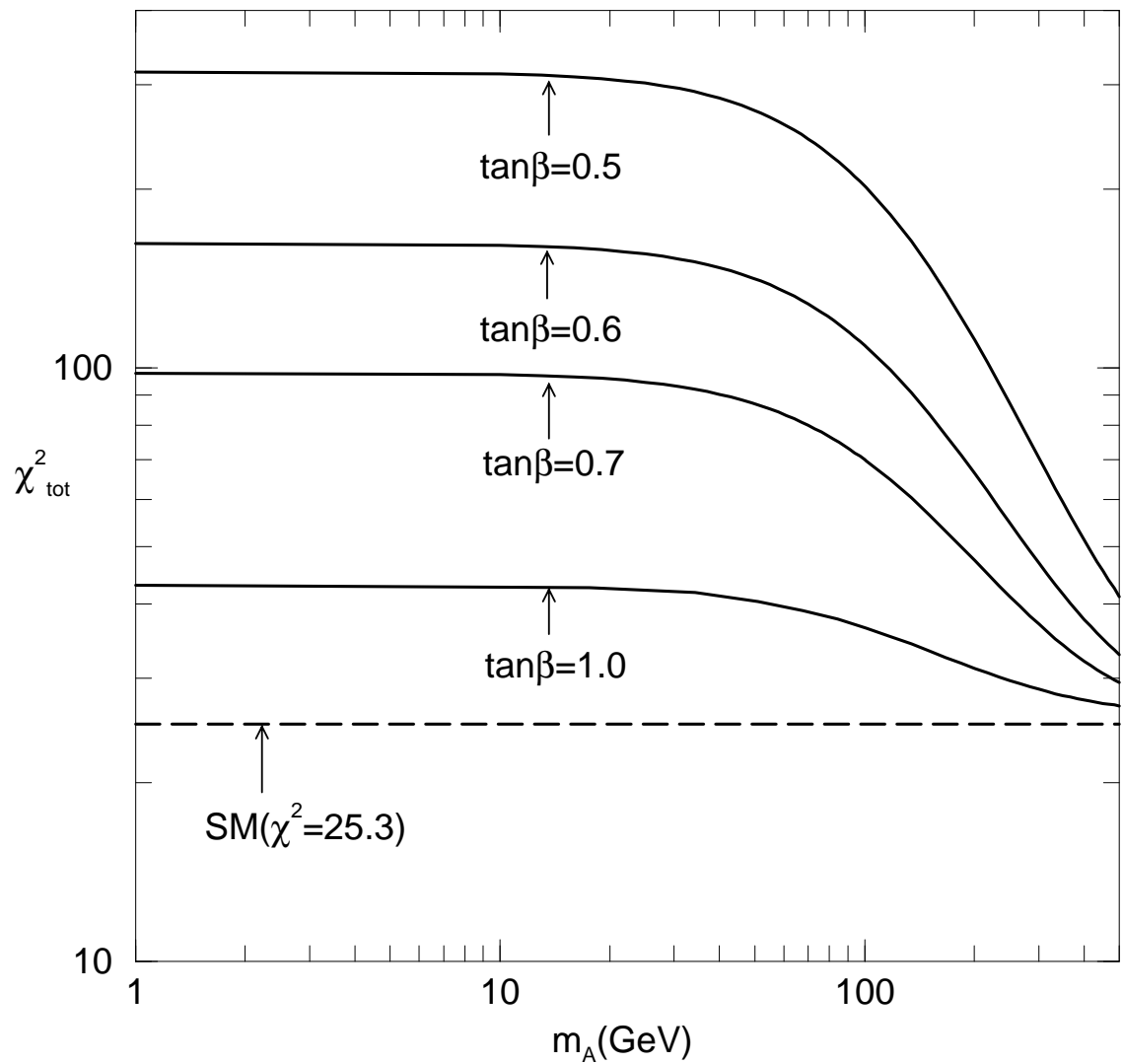}}
}
\smallskip\smallskip\smallskip\smallskip
\caption{The total $\chi^2$ as a function of $m_A$ for various 
values of $\tan\beta$. The four thick lines correspond to 
$\tan\beta=0.5$, 0.6, 0.7 and 1.
The thick horizontal dashed line denotes the minimum $\chi^2$ in the 
SM ($\chi^2=25.3$). 
}
\end{figure}


\end{document}